\documentclass[showpacs,twocolumn,preprintnumbers,amsmath,amssymb]{revtex4}
\usepackage{epsfig}

\begin{document}

\title{Modeling of microarray data with zippering}

\author{J. M. Deutsch$^1$, Shoudan Liang$^2$ and Onuttom Narayan$^1$}
\affiliation{
$^1$Department of Physics, University of California, Santa Cruz, CA 95064.\\
$^2$  National Aeronautics and Space Administration Advanced Supercomputing
   Division, National Aeronautics and Space Administration Ames Research Center,
Moffet Field, CA 94035
}

\date{\today}

\begin{abstract}
The ability of oligonucleotide microarrays to measure gene expression has been
hindered by an imperfect understanding of the relationship between input
RNA concentrations and output signals. We argue that this relationship
can be understood based on the underlying statistical mechanics of these
devices. We present a model that includes the relevant interactions
between the molecules. Our model for the first time accounts for
partially zippered probe-target hybrids in a physically realistic manner, and
also includes target-target binding in solution.
Large segments of the target molecules are not bound to the probes,
often in an asymmetric pattern, 
emphasizing the importance of modeling zippering properly.  
The resultant fit between the
model and training data using optimized parameters is excellent, and it also
does well at predicting test data.
\end{abstract}


\maketitle 

Oligonucleotide microarrays have had a profound impact
on medical diagnosis and molecular biology. These devices have thousands of
cells, each containing numerous copies
of a specific DNA probe attached to the substrate. After amplification, cRNA (or DNA) derived from a biological sample 
is fluorescently labeled and then fragmented into targets that are allowed to bind
to the probes. The targets hybridize with their complementary DNA
probes with high specificity. The fluorescence allows
an optical readout of the concentration of thousands of RNA transcripts 
simultaneously.

There are often orders of magnitude difference in
the concentrations of the various kinds of targets
and many important genes are expressed only in very small
concentrations. These can be hard to measure by this method for reasons
we now discuss. 

In Affymetrix GeneChips, 16 kinds of probe, all with the same concentration and
probing different segments of a single mRNA, are employed for every transcript.
There
can be {\em order of magnitude variations} in the signal intensities between
these 16 probes. The specific  signal intensities are reproducible and so 
cannot be explained as statistical error. Instead, they occur because
the interactions between the probes and the target molecules are complex.
In addition to the binding between a probe and its complementary target
(specific binding), there are binding of targets to noncomplementary probes 
(nonspecific binding) and target-target binding. In
addition, even when a target binds to its complementary probe,
it is possible that it is only partially hybridized, with the ends
unzipped. This is likely to be an important effect in microarrays, since
their operating temperature is close to the melting temperature of the
hybrids, so that substantial fluctuations in binding can be expected.

Various statistical techniques have been used to reduce the errors
caused by these effects. 
For example Affymetrix, in addition to using 16 probes for every transcript,
has added another set of ``mismatch" probes that differ from the original
``perfect match"
probes by the alteration of the middle nucleotide. These are compared
to reduce the error from nonspecific binding. 
Although these
techniques reduce the uncertainties in predicted input concentrations,
by analyzing 32 numbers,
they still have difficulty in detecting small concentrations of RNA
at biologically significant levels. Alternative statistical approaches
have been proposed~\cite{liwong}.

As recognized by several previous authors~\cite{held,hekstra,zhang}, 
the statistical techniques used
by Affymetrix and others do not utilize the probe sequence information.
The hope is that including these should greatly increase the reliability of predictions.

Held et al~\cite{held} modeled the binding of each target molecule to
its complementary probe using a Langmuir adsorption model~\cite{atkins},
If $c$ is the concentration in solution of the
target,
the fraction of
probe molecules that are hybridized is
\begin{equation}
f_h = {1\over{1 + \exp[\beta(\Delta G-\mu)]}}
\label{fh}
\end{equation}
where $ \exp(\beta\mu)\propto c$.
The binding
energy $\Delta G$~\cite{foot_wrong_F} was obtained from previously measured stacking free
energies~\cite{santalucia,turner}. 
All target-probe pairs were treated independently,
with nonspecific binding included phenomenologically by adding a 
$\Delta G$ dependent 
constant to the measured signal. 
to be of the form $a + b \exp[ c \Delta G].$ 
Partially hybridized probes were not considered, although the authors
comment that they might be of importance. Perhaps because of 
this, in their Boltzmann factors $\sim\exp(\beta\Delta
G),$ the best fit effective temperature was approximately seven times
the actual temperature. 

Hekstra et al~\cite{hekstra} also used a Langmuir adsorption model with
an additive background from nonspecific binding. However, the resultant
three parameters for each probe-target pair were not evaluated using
previously determined
stacking energies, but were fitted to linear combinations of the number 
of each nucleotide.

Zhang et al~\cite{zhang} attempted to include the effects of partial binding 
of target molecules to probes. The model they came up with was the ``positional
dependent nearest neighbor model" (PDNN) where the binding energy for a probe-target pair
was taken to be of the form $\Delta G = \sum_k \omega_k \epsilon(b_k, b_{k+1}),$
where $\epsilon$ is the stacking energy for the adjacent bases $b_k$ and 
$b_{k + 1},$ and $\{\omega_k\}$ was a set of weights depending only on
the position on the probe $k$, and not on the specific probe molecule. 
The 
probe intensities were assumed to be linear in the target concentrations,
i.e. saturation effects were not included. 

The above methods incorporating sequence information have
different approaches to data fitting, with various
physical effects included. 
In this paper we present a comprehensive approach that includes what
we believe to be all the most
important effects, to construct a physical model for microarrays.
In particular we have included ``zippering effects", i.e. target molecules
partially hybridized to probes, by using a full statistical mechanical approach
rather than ad-hoc position-dependent weights. This reduces the number of
fitting parameters enormously, because partial binding can be understood
completely as a consequence of the stacking energies. We also include
the effect of target-target binding in solution. As pointed out by
Held et al~\cite{held}, the saturation intensities of different probes
that correspond to target fragments from the same mRNA molecule can be
different by an order of magnitude. It is likely that this is because of
these target target interactions, reducing the effective concentration
in solution of different targets by different amounts. Since
RNA-RNA interactions are stronger than RNA-DNA~\cite{sugimoto}, this effect can be substantial. We also include non-specific binding for each probe similarly.

We present our model in three parts: specific binding, including zippering, 
non-specific binding, and finally target-target interactions in solution. 

Consider a target molecule consisting of a sequence of bases $\{b_1\ldots
b_N\}$ (where $N = 25$ for the Affymetrix GeneChip). According to the
stacking energy description of hybridization, if this target molecule
is fully bound to its complementary probe, the resultant change in free
energy is of the form
\begin{equation}
\Delta G(1, N) = \sum_{k=1}^{N-1} \epsilon(b_k, b_{k+1})+\epsilon_i,
\label{stacking}
\end{equation}
where $\epsilon_i$ is the initiation energy of attachment.  However,
it is also possible for the target to be partially bound to the probe,
with only the bases $n$ to $m$ being bound. This configuration would have
a $\Delta G(n, m)$ given by Eq.(\ref{stacking}) but with the sum from $k=n$
to $k=m-1.$ Because the local stiffness is large, and the target-probe
hybrid is in a helical structure, we only need to consider
configurations for which the unbound parts start at the ends, rather
than forming isolated islands in the middle. Thus a target-probe pair
can be viewed as a double-ended zipper.

The resultant partition function for the bound state that includes all 
partially bound configurations is 
\begin{equation}
Z = \sum_{n< m} \exp(-\beta\Delta G(n, m)) \equiv \exp(-\beta \Delta G)
\label{partfn}
\end{equation}
where $\Delta G$ is the total binding free energy.
Naively, this takes $O(N^3)$ operations to compute, which is prohibitively
expensive, because $\Delta G$ is computed repeatedly when optimizing the model.
However, $Z$ can be computed in $O(N)$ operations using  recursion
relations.  
Define $Z(i)$ as the analog of $Z$ in Eq.(\ref{partfn}),
but with only the bases from 1 to $i$ included. $Z(i)$ is the sum of
two terms: $Z_u(i),$ which considers configurations that are unbound
at the site $i$ (but have a bound segment somewhere before $i$), and
$Z_b(i),$ which considers configurations that are bound at the site $i.$
The recursion relations for these are
\begin{eqnarray}
Z_u(i+1) & = & Z_u(i) + Z_b(i) \nonumber\\
Z_b(i+1) & = & Z_b(i)\exp(-\beta\epsilon(b_i, b_{i+1})) + \exp(-\beta\epsilon_i).
\label{recurs}
\end{eqnarray}
\vskip 0.2in
In the first of these equations, configurations that are attached at $i$
can detach at $i+1.$ However, in the second equation, because we allow
only one bound segment in our zipper model, and configurations in $Z_u(i)$
have already had a bound segment before the base $i,$ there is no contribution
from $Z_u.$ 
These recursion relations are 
are a simplification of those 
in Ref.~\cite{orland,poland}. 
The $\Delta G$ from Eq.(\ref{partfn}) can now be used in Eq.(\ref{fh}).

The next effect we consider is nonspecific binding. In
principle, this could be accomplished with an approach very similar to
the one we have constructed for specific binding. This would require a
complete knowledge of all molecular fragments present in the solution;
these include a background of human RNA, and the additional ``spiked-in"
target molecules that the experiment tries to measure.  It would also
be necessary to determine the stacking energies for all $4^4$ possible
mismatched (or matched) sequences of two base pairs. Since neither 
of these
is fully known, we use a
statistical approach to model non-specific binding. We assume that the RNA
giving rise to non-specific binding is sufficiently diverse that it can be
treated as a `bath' of random sequences. We also make the approximation
that stacking energies can be defined in terms of the nearest neighbors
on the probe. If $\{b_k\}$ is the probe sequence and $\{c_k\}$ is a 
(non-complementary) target sequence, this approximation is
\begin{equation}
\sum_{\{c_k\}} \prod_k e^{-\beta\epsilon(b_k, b_{k+1}; c_k, c_{k+1})} =
\prod_k e^{-\beta\epsilon^\prime(b_k, b_{k+1})}
\end{equation}
where $\epsilon^\prime$ is an effective stacking energy. In the absence
of experimental knowledge of the 256 stacking energies, not using
this approximation would result in so many adjustable parameters in our
model so as to make any results meaningless.

Finally, we include target-target binding in solution. 
As mentioned earlier,
there are substantial variation in the saturated probe intensities
for different target fragments from a single mRNA molecule. Further,
the intensities saturate at a much higher concentration of target
molecules than one would expect from the number of probe molecules in a single
cell~\cite{foot_saturate}. This strongly suggests that the target
concentration in solution is significantly depleted, by an amount
that differs from one target species to another. 
We model the fraction
of any target species that is lost to target-target binding by using
the analogs of Eq.(\ref{fh}).
For reasons similar to those for non-specific binding~\cite{foot_rna_rna},
we model this statistically. 

We performed numerical simulations using the model described 
above using Affymetrix's
Series 1532 Latin Square~\cite{irizarry} data. In these experiments, a cocktail
of human cRNA was ``spiked in" to a background of human RNA of unknown
composition. In any experiment, the transcripts had concentrations $0,
0.25, 0.5\ldots 1024$ pM. The transcripts whose concentration was large
varied from one experiment to another cyclically, in a pattern that formed
a Latin Square matrix. 
Each cRNA transcript could hybridize with 16 different probe sequences 
at different positions along the original transcript.

In the simulations, we varied the parameters of the model in order to
minimize the fitness, the log mean square difference between the measured
signal intensities and the model predictions:
\begin{equation}
F = \sum [\ln I_{meas} - \ln I_{pred}]^2/M
\label{fitness}
\end{equation}
where the sum runs over probes and experiments, and M is the total
number of data points used. We followed the common practice of using the
logarithm in this definition, because the intensities vary over orders
of magnitude, and doing otherwise would discount low concentration
transcripts, which are important biologically. The data from probes
407\_at and 36889\_at was not used, following Affymetrix's recommendation. The 
minimization of $F$ as a function of model parameters is a computationally
intensive
optimization problem. In order to 
increase convergence to the solution, we used several techniques, of 
which parallel tempering~\cite{parallel} was found to work best.

The parameters in the model were 16 stacking energies and one
initiation energy for specific, non-specific and target-target binding
(51 parameters), the number of probe molecules for each species,
a scale factor converting from hybridization to signal intensity,
and a (small) uniform background to the signal. The total number of
parameters was 54. Although DNA-RNA stacking energies for matched bases
are known~\cite{sugimoto}, we allowed these to vary as free parameters.
This was because the addition of fluorescent tags to the RNA molecules
has been shown~\cite{rockefeller}
to change the stacking energies. The final optimized
stacking energies were of the same order of magnitude as experimental
results for untagged molecules~\cite{sugimoto}.

Although 54 fitting parameters might seem to be a large number, the model
was used to fit 2464 data points, and our number of parameters compares
favorably with prior work~\cite{zhang}. Importantly, when we randomly
shuffled the sequences associated with the different probes and redid 
the optimization, the function
$F$ in Eq.(\ref{fitness}) increased by a factor of more than two, indicating that
our results were not an artifact of having too many parameters.

Apart from being able to fit the given data accurately, the purpose
of a model is to be able to analyze new experiments with
unknown concentrations of transcripts, and accurately predict these
concentrations. Accordingly, we left out some some (one or three) of
the transcripts during parameter estimation (in addition to 
transcripts 407\_at and 36889\_at mentioned earlier). After this data
had been used to train the model, we tried to predict the
concentrations of the transcripts that had been left out.

\begin{figure}
\centerline{\epsfxsize=\columnwidth \epsfbox{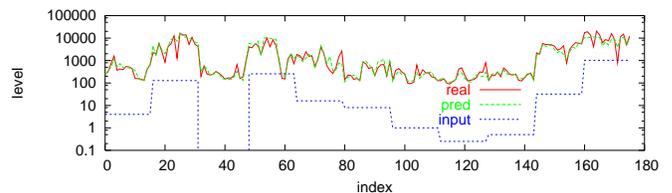}}
\caption{Plot of the signal intensities as measured (solid line)
and from the model (dashed line) for all the probes (except for 
those corresponding to three transcripts) for a typical
experiment.  The input concentration of each
transcript is also shown (not on the same scale); since there are
sixteen probes for each transcript, this is a piecewise constant curve.
}
\label{intensity_vs_probe}
\end{figure}

Figure~\ref{intensity_vs_probe} shows the measured signal intensities for all the probes
in a typical experiment, except for those corresponding to three 
transcripts: one to be used in 
the subsequent prediction stage, and the two that were 
known to be unreliable.  The figure also shows the
signal intensities from the model, with optimized parameters, and the
input concentrations. The model reproduces the measured intensities quite
well. The model parameters were optimized
once for all the experiments simultaneously.  The residual error $F,$
defined in Eq.(\ref{fitness}), was 0.19.

\begin{figure}
\centerline{\epsfxsize=\columnwidth \epsfbox{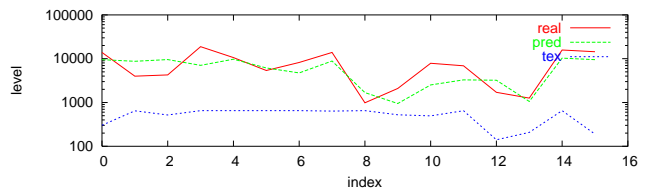}}
\caption{Measured (solid line) and predicted (dashed line) intensities
for probes for transcript 37777\_at from the Affymetrix Latin Square
data. This transcript was not used for model optimization. 
Results of a similar procedure with the 
``PDNN" model~\cite{zhang} are also shown (dotted line).  
}
\label{pred_vs_probe}
\end{figure}

With the optimal parameters
obtained above, the model was used to predict the signal intensities
for the probes corresponding to the transcript that was left out in the
training stage, as shown in Figure~\ref{pred_vs_probe}. The same
figure shows results (with the same procedure) using
the PDNN model of Ref.~\cite{zhang}.  As seen in the figure, the predictions
with our model are much better.

\begin{figure}[t]
\epsfxsize=\columnwidth \epsfbox{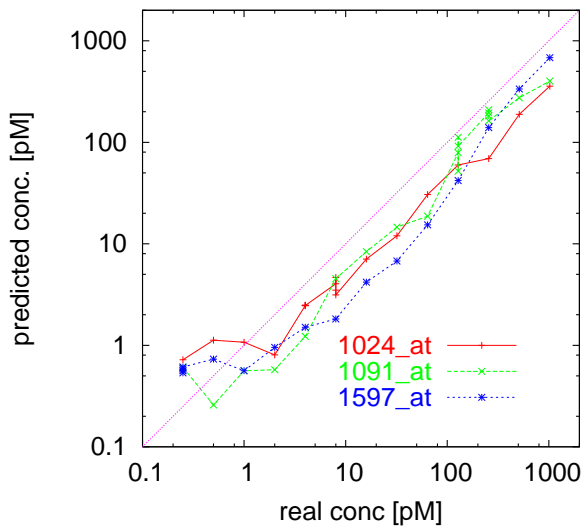}
\caption{Predicted input concentrations as a function of the actual
input concentrations for three transcripts. These were not included in
the parameter optimization of the model.
}
\label{pred_vs_conc}
\end{figure}

We also show the prediction capabilities of the model with a different
procedure, which is similar to the one used by Ref.~\cite{hekstra}. Three
transcripts were left out (in addition to the two faulty ones) in
the training stage. In the prediction stage, for each experiment, the
sixteen measured probe intensities were used collectively to predict
the input concentration of each of the three excluded 
transcripts~\cite{foot_pred_details}. The results are shown 
in Figure~\ref{pred_vs_conc}.

When target-target interactions were omitted from our model,
the residual error $F$ increased to
0.27. However, the significance of this is hard to interpret,
since the number of parameters is thereby reduced from 54 to 37.
Therefore, we assessed this reduced model for prediction. 
We found 
a noticeable degradation in predictive power at low input
concentrations compared to Figure~\ref{pred_vs_conc}.

\begin{figure}[htb]

{\epsfxsize=\columnwidth \epsfbox{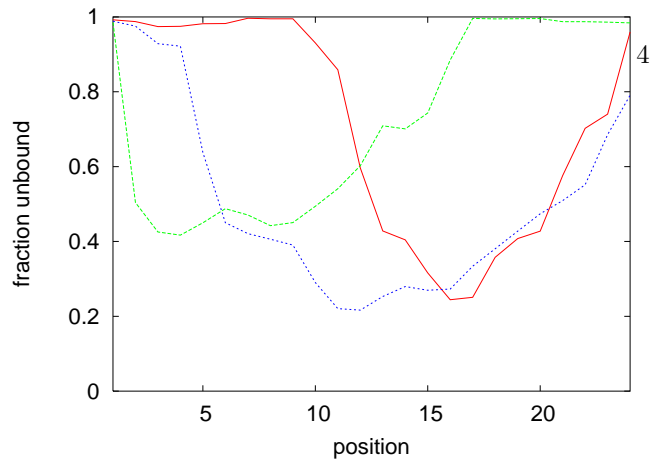}}
\caption{Fraction of time a base is unbound
as a function of position (from the solution end) along the probe,
for 3 different
probes.
}
\label{partial_binding}
\end{figure}

Lastly, in Figure~\ref{partial_binding} we present some data on partial 
zippering. The figure 
displays the fraction of the time a base pair is unbound as a function of
position on the probe. As is evident, these probes are only partially bound.
The places they bind are not symmetric about the middle, and depend
strongly on the probe sequence.  This kind of behavior differs from that of
previous models.  Evidence for this kind of asymmetric
partial binding can be seen from careful experiments on Agilent
microarrays~\cite{hughes}. 

In this paper, we have constructed a physical model for hybridization of
RNA transcript targets to probes in microarrays, in order to predict 
experimental signal intensities. Our model 
includes for the first time the effect of partial hybridization of probes
(zippering) derived from fundamental statistical mechanics,
and also the effect of target-target interactions. The prediction
capabilities of our model appear to surpass those of other approaches.


\begin{references}
\bibitem{foot_wrong_F} Although DNA-DNA stacking energies appear to have 
been used instead of the appropriate DNA-RNA~\cite{sugimoto}.
\bibitem{foot_saturate} The number of probe molecules can be estimated from
their density~\cite{southern}, about 1 molecule every 40
square $\AA,$ 
and the solution volume (200 $\mu L$). This gives $\approx 0.2 pM$.
\bibitem{foot_rna_rna} We have verified that all the mismatched 
RNA-RNA stacking energies are not known, unlike often believed.
\bibitem{foot_pred_details} After optimization, the nonlinear
relation between input concentration and output signal was inverted. 
When the measured signal exceeded the saturation value, a large
positive value was arbitrarily assigned. The training data was run
through this inversion procedure to create a calibrated prediction of
input concentration. This calibration function was used on the test data.
\end{references}
\end{document}